\documentclass[a4paper,english,aps, prl, twocolumn]{revtex4-1}
\usepackage[T1]{fontenc}
\usepackage[latin9]{inputenc}
\usepackage{babel}
\usepackage{array}
\usepackage{multirow}
\usepackage{amsmath}
\usepackage{amssymb}
\usepackage{graphicx}
\usepackage{wasysym}
\usepackage[unicode=true,pdfusetitle,
 bookmarks=true,bookmarksnumbered=false,bookmarksopen=false,
 breaklinks=false,pdfborder={0 0 1},backref=false,colorlinks=false]
 {hyperref}

\makeatletter

\providecommand{\tabularnewline}{\\}

\makeatother

\begin{document}
\title{The role of flexural coupling in heat dissipation from a two-dimensional
layered material to its hexagonal boron nitride substrate}
\author{Zhun-Yong Ong}
\email{ongzy@ihpc.a-star.edu.sg}

\affiliation{Institute of High Performance Computing, A{*}STAR, Singapore 138632,
Singapore}
\author{Gang Zhang}
\affiliation{Institute of High Performance Computing, A{*}STAR, Singapore 138632,
Singapore}
\author{Yong-Wei Zhang}
\affiliation{Institute of High Performance Computing, A{*}STAR, Singapore 138632,
Singapore}
\date{\today}
\begin{abstract}
Understanding the limits of phononic heat dissipation from a two-dimensional
layered material (2DLM) to its hexagonal boron nitride (h-BN) substrate
and how it varies with the structure of the 2DLM is important for
the design and thermal management of h-BN-supported nanoelectronic
devices. We formulate an elasticity-based theory to model the phonon-mediated
heat dissipation between a 2DLM and its h-BN substrate. By treating
the h-BN substrate as a semi-infinite stack of harmonically coupled
thin plates, we obtain semi-analytical expressions for the thermal
boundary conductance (TBC) and interfacial phonon transmission spectrum.
We evaluate the temperature-dependent TBC of the $N$-layer 2DLM (graphene
or MoS$_{2}$) on different common substrates (h-BN vs. a-SiO$_{2}$)
at different values of $N$. The results suggest that h-BN is substantially
more effective for heat dissipation from MoS$_{2}$ than a-SiO$_{2}$
especially at large $N$. To understand the limitations of the our
stack model, we also compare its predictions in the $N=\infty$ limit
to those of the more exact Atomistic Green's Function model for the
graphite-BN and molybdenite-BN interfaces. Our stack model provides
clear insights into the key role of the flexural modes in the TBC
and how the anisotropic elastic properties of h-BN affect heat dissipation.
\end{abstract}
\maketitle

\section{Introduction}

Bulk hexagonal boron nitride (h-BN) is widely used as a substrate
and encapsulating material for nanoscale devices based on two-dimensional
(2D) layered materials~\citep{CDean:NatNano10_BoronNitride,GLee:ACSNano13_Flexible}
because its atomically flat surface, which is free of dangling bonds
and charged impurities, enhances device performance by reducing the
mechanical perturbation to the active 2D layered material (e.g. graphene
or MoS$_{2}$), unlike conventional insulating substrates, such as
amorphous SiO$_{2}$ (a-SiO$_{2}$) and Al$_{2}$O$_{3}$, which have
corrugated surfaces~\citep{WCullen:PRL10_High,EPaek:JAP13_Computational,ZYOng:JAP18_Flexural,JGuo:JPDAP19_Conformal}.
In addition, the h-BN has other material properties (e.g. electrically
insulating, chemical stability, mechanical flexibility and high in-plane
thermal conductivity~\citep{LLindsay:PRB11_Enhanced,LLi:AFM16_Atomically,AFalin:NatCommun17_Mechanical,JWangLRSCAdv17_Graphene})
which are highly advantageous for the development of 2D nanoelectronics.

In spite of its many desirable material properties, a potential obstacle
to the integration of h-BN into 2D nanoelectronic devices is the thermal
resistance of its interface with the two-dimensional layered material
(2DLM), which can lead to a heat dissipation bottleneck and limit
device performance if the Joule heat is not removed efficiently~\citep{EPop:NR10_Energy}.
In an active 2DLM-based device on a h-BN substrate, the generated
Joule heat is transferred from the 2DLM to the h-BN across their interface
through the relatively weak van der Waals (vdW) forces coupling of
the thermal motion of the atoms in the 2DLM and on the substrate surface
~\citep{ZYOng:2DM19_Energy,DRhodes:NatMater19_Disorder}. The rate
at which heat is dissipated across this interface varies with its
thermal boundary conductance (TBC) and depends on the strength of
the vdW forces and the elastic properties of the 2DLM and h-BN.

In bulk h-BN, the stacking of the individual h-BN layers, which is
responsible for its characteristic surface flatness, also gives rise
to its anisotropic elastic properties and, in a 2D device, affects
how the vdW forces dissipate energy across the interface between the
2DLM and its h-BN substrate. Given the potential of h-BN for 2D device
applications, physical insights into how a 2DLM dissipates heat through
its planar interface with h-BN are critical for the development of
thermally efficient nanoelectronics~\citep{EPop:NR10_Energy,ZYOng:2DM19_Energy}
as well as for understanding the theoretical limits of substrate-directed
heat dissipation. Although atomistic thermal transport simulations
can play an important role in obtaining these insights~\citep{YLiu:SciRep17_Thermal},
they sufer from finite-size effects and it is more challenging to
connect their results to the elastic properties of the materials.
Hence, it is useful to develop theoretical models that use simple
elastic parameters (e.g. bending rigidity and interlayer force constants)
as inputs because they can be modified easily to predict how the choice
of substrate material and device structure (e.g. dielectric encapsulation
and thickness of the 2DLM) affects heat dissipation or isolation~\citep{SVaziri:SciAdv19_Ultrahigh}.
Such models have been developed for isotropic elastic solid substrates,
such as a-SiO$_{2}$~\citep{BNJPersson:JPCM11_Phononic,ZYOng:PRB16_Theory},
to understand the effects of encapsulation~\citep{ZYOng:PRB16_Theory}
and the thickness of the 2DLM~\citep{ZYOng:PRB17_Thickness} on the
TBC but they cannot be applied to h-BN at present because of its highly
anisotropic elastic properties.

To understand how the elastic properties of the substrate affect heat
dissipation, let us recall the formula for the phonon TBC which we
can write in the Landauer form~\citep{ZYOng:PRB16_Theory} as 
\begin{equation}
G_{\text{ph}}=\int\frac{d\omega}{2\pi}\hbar\omega\frac{dN(\omega,T)}{dT}\xi(\omega)\ ,\label{eq:LandauerTBCFormula}
\end{equation}
where $N(\omega,T)$ denotes the Bose-Einstein distribution function,
\begin{equation}
\xi(\omega)=\int_{q<q_{c}}\frac{d^{2}q}{(2\pi)^{2}}\Xi(\boldsymbol{q},\omega)\label{eq:TotalTransmission}
\end{equation}
is the total areal transmission function at frequency $\omega$, and
$q_{c}$ is the cutoff transverse wave vector of the 2DLM such that
$\pi q_{c}^{2}$ is the area of its Brillouin zone. In Eq.~(\ref{eq:TotalTransmission}),
$\xi(\omega)$ can be interpreted as the spectral window for phonon
transmission and is obtained by summing the modal transmission function
$\Xi(\boldsymbol{q},\omega)$ over the 2D transverse wave vector $\boldsymbol{q}$.
The expression for $\Xi(\boldsymbol{q},\omega)$ is~\citep{ZYOng:PRB16_Theory}
\begin{equation}
\Xi(\boldsymbol{q},\omega)=\frac{4K^{2}\text{Im}D_{\text{sub}}(\boldsymbol{q},\omega)\text{Im}D_{\text{2D}}(\boldsymbol{q},\omega)}{|1-K[D_{\text{sub}}(\boldsymbol{q},\omega)+D_{\text{2D}}(\boldsymbol{q},\omega)]|^{2}}\ ,\label{eq:ModalTransmission}
\end{equation}
where $K$ is the areal spring constant at the 2DLM-substrate interface,
$D_{\text{sub}}(\boldsymbol{q},\omega)$ is the Green's function for
the substrate surface displacement which describes how it responds
to an applied normal force, and $D_{\text{2D}}(\boldsymbol{q},\omega)$
is the Green's function for the flexural motion of the 2DLM which
can be single or multilayered. Equation~(\ref{eq:ModalTransmission})
satisfies the condition $0\leq\Xi(\boldsymbol{q},\omega)<1$~\citep{ZYOng:PRB16_Theory}
and can be interpreted as the fraction of energy transmitted across
the 2DLM-substrate interface at $\boldsymbol{q}$ and $\omega$. It
also assumes that the energy transfer at the interface is due entirely
to the harmonic coupling between the 2DLM and its substrate, with
the anharmonicity of the interfacial bonds playing an insignificant
role. In this model~\citep{ZYOng:PRB16_Theory}, we assume that the
$D_{\text{sub}}(\boldsymbol{q},\omega)$ and $D_{\text{2D}}(\boldsymbol{q},\omega)$
are isotropic with respect to $\boldsymbol{q}$, depending only on
its magnitude $q=|\boldsymbol{q}|$ and $\omega$. The analytical
expression for the $D_{\text{sub}}(\boldsymbol{q},\omega)$ of an
\emph{isotropic} elastic solid substrate, which depends on the speed
of sound and mass density of the solid, is derived in Ref.~\citep{BPersson:JCP01_Theory}
and has been used with Eq.~(\ref{eq:LandauerTBCFormula}) to estimate
the TBC between 2DLMs (e.g. graphene and MoS$_{2}$) and a-SiO$_{2}$
~\citep{BNJPersson:JPCM11_Phononic,ZYOng:PRB16_Theory,ZYOng:PRB17_Thickness}.
However, for a layered substrate such as h-BN, our ability to calculate
the TBC of the 2DLM-substrate interface is limited by the lack of
an expression for $D_{\text{sub}}(\boldsymbol{q},\omega)$ that describes
its anisotropic elastic character and layered geometry.

In this paper, we address this problem by formulating an anisotropic
model of the h-BN substrate that is based on linear elasticity theory
and treats the h-BN lattice as a semi-infinite stack of harmonically
coupled thin plates. Our approach builds on the theoretical concepts
developed in Refs.~\citep{ZYOng:PRB16_Theory,ZYOng:PRB17_Thickness}.
Although our `stack model' is specifically used to treat h-BN in this
paper, the theory can be easily extended to other 2D layered analogs
of bulk h-BN (e.g. graphite) and may also be useful for understanding
the dynamics of breathing modes in 2DLMs~\citep{YZhao:NL13_Interlayer,LLiang:ACSNano17_LowFrequency}
as well as for estimating the substrate-induced changes in the properties
of a h-BN-supported 2DLM using many-body techniques~\citep{BAmorim:PRB13_Flexural}.
We solve our stack model to derive an analytical expression for $D_{\text{sub}}(\boldsymbol{q},\omega)$
which is verified numerically. We then apply Eq.~(\ref{eq:LandauerTBCFormula})
to compute the TBC for different 2DLMs (graphene and MoS$_{2}$) and
substrates (a-SiO$_{2}$ and h-BN). We analyze how the substrate affects
the dependence of the TBC on the temperature and number of layers
in the 2DLM. We also compare the predictions of our stack model to
those of the Atomistic Green's Function (AGF) method~\citep{YLiu:SciRep17_Thermal},
which is computationally more expensive and atomistically more detailed,
to uncover the role of the flexural phonons in interfacial thermal
transport. Finally, we examine how the TBC varies with the interlayer
spring constant strength from the weak (van der Waals) to the strong
coupling limit. We note here that the model described in this paper
can be extended to 2DLMs with defects, such as grain boundaries, by
modifying the bending rigidity and flexural phonon lifetimes which
are reduced by defect scattering.~\citep{CAPolanco:PRB18_AbInitio}

\section{Methodology}

\subsection{Theory of h-BN}

\subsubsection{Stack model of harmonically coupled thin plates}

\begin{figure}
\begin{centering}
\includegraphics[width=7cm]{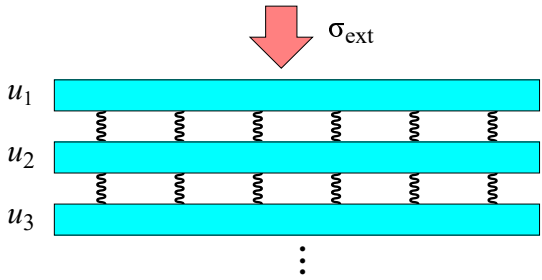}
\par\end{centering}
\caption{Schematic of the semi-infinite stack model for h-BN. The external
normal force $\sigma_{\text{ext}}$ is applied to layer 1, which is
at the top of the stack, while the flexural motion of the individual
layers is coupled harmonically via spring-like harmonic forces as
indicated by the wavy lines.}
\label{fig:hBNGeometry}
\end{figure}

The key idea in our derivation of $D_{\text{sub}}$ is that the h-BN
lattice is treated as a continuum in the in-plane $x$ and $y$ directions
and as a discrete system in the cross-plane $z$ direction in order
to capture the anisotropic character of the lattice. As in Refs.~\citep{ZYOng:PRB16_Theory}
and \citep{ZYOng:PRB17_Thickness}, we only consider the flexural
motion and ignore the in-plane polarized degrees of freedom in the
individual layers for the sake of simplicity. We treat the h-BN substrate
as a semi-infinite stack of sheets, indexed $n=1,\ldots,\infty$ where
layer 1 is at the top surface, as shown in Fig.~\ref{fig:hBNGeometry}
with each sheet is modeled as a thin plate. We assume that adjacent
sheets, separated by an interlayer distance of $a$, are coupled via
a harmonic force. The stacking order of the individual h-BN layers
is ignored in our model because of its continuum nature.

Our objective is to determine how the flexural displacement of the
top layer $u_{1}(\boldsymbol{r},t)$, where $\boldsymbol{r}=(x,y)$
and $t$ represent respectively the in-plane position and time, oscillates
when an external harmonic normal force $\sigma_{\text{ext}}(\boldsymbol{r},t)$
is applied on layer 1 in the $z$ direction. In other words, we want
to find the frequency response function $D_{11}(\boldsymbol{q},\omega)$,
which is the $D_{\text{sub}}(\boldsymbol{q},\omega)$ in Eq.~(\ref{eq:ModalTransmission}),
in the linear relationship
\begin{equation}
\tilde{u}_{1}(\boldsymbol{q},\omega)=-D_{11}(\boldsymbol{q},\omega)\tilde{\sigma}_{\text{ext}}(\boldsymbol{q},\omega)\label{eq:Flexural_linear_response}
\end{equation}
where $\tilde{u}_{1}(\boldsymbol{q},\omega)=\frac{1}{(2\pi)^{3}}\int d\boldsymbol{r}\int dtu_{1}(\boldsymbol{r},t)e^{-i(\boldsymbol{q}\cdot\boldsymbol{r}-\omega t)}$
is the Fourier transform of the position and time-dependent flexural
displacement of layer $1$ $u_{1}(\boldsymbol{r},t)$, and $\tilde{\sigma}_{\text{ext}}(\boldsymbol{q},\omega)=\frac{1}{(2\pi)^{3}}\int d\boldsymbol{r}\int dt\sigma_{\text{ext}}(\boldsymbol{r},t)e^{-i(\boldsymbol{q}\cdot\boldsymbol{r}-\omega t)}$
is the Fourier transform of the position and time-dependent applied
external normal force $\sigma_{\text{ext}}(\boldsymbol{r},t)$ . Finding
this response of the top layer is however complicated by the collectively
coupled motion of the layers in the semi-infinite stack. In the rest
of the subsection, we show how an analytical expression for $D_{11}(\boldsymbol{q},\omega)$
can be obtained by exploiting the matrix structure of the coupled
equations of motion. To avoid confusion with the symbol for angular
frequency $\omega$ in the rest of the paper, we use the symbol $u$
instead of $w$ as per the convention in linear elasticity theory
to represent displacement in the out-of-plane $z$ direction.

We begin our derivation with the equations of motion for the individual
layers in h-BN. The equation of motion for the top layer (layer 1)
is 
\begin{equation}
\rho\frac{\partial^{2}u_{1}}{\partial t^{2}}=-\kappa\nabla^{2}\nabla^{2}u_{1}-g(u_{1}-u_{2})+\sigma_{\text{ext}}\ ,\label{eq:EOM_Sheet1}
\end{equation}
where $\nabla^{2}=\frac{\partial^{2}}{\partial x^{2}}+\frac{\partial^{2}}{\partial y^{2}}$
is the in-plane Laplace operator, $\rho$ is the areal mass density,
$u_{n}=u_{n}(\boldsymbol{r},t)$ is the position and time-dependent
flexural displacement of layer $n$ ($n=1,\ldots,\infty$) from its
equilibrium position, $\kappa$ is the sheet bending rigidity, and
$g$ is the interlayer areal spring constant. The second term on the
righthand side (RHS) of Eq.~(\ref{eq:EOM_Sheet1}) represents the
harmonic coupling between layers 1 and 2. If we set $g=0$ and $\sigma_{\text{ext}}=0$,
we recover the classic Kirchhoff-Love equation~\citep{YWei:NSR19_Nanomechanics}
($\rho\frac{\partial^{2}u_{1}}{\partial t^{2}}=-\kappa\nabla^{2}\nabla^{2}u_{1}$)
for the flexural motion of a thin plate~\citep{BNJPersson:JPCM11_Phononic,ZYOng:PRB16_Theory}.
The equation of motion for the remaining subsurface layers ($n\geq2$)
in general can be similarly written as
\begin{equation}
\rho\frac{\partial^{2}u_{n}}{\partial t^{2}}=-\kappa\nabla^{2}\nabla^{2}u_{n}-g(u_{n}-u_{n-1})-g(u_{n}-u_{n+1})\ ,\label{eq:EOM_Sheetn}
\end{equation}
where the second and third terms on the RHS of Eq.~(\ref{eq:EOM_Sheetn})
represent the harmonic coupling to its top and bottom neighboring
layers, respectively. If we define the Fourier transform of a function
$f(\boldsymbol{r},t)$ as $\tilde{f}(\boldsymbol{q},\omega)=\frac{1}{(2\pi)^{3}}\int d\boldsymbol{r}\int dtf(\boldsymbol{r},t)e^{-i(\boldsymbol{q}\cdot\boldsymbol{r}-\omega t)}$,
then the Fourier transform of Eq.~(\ref{eq:EOM_Sheet1}) yields the
algebraic equation
\begin{equation}
-\rho\omega^{2}\tilde{u}_{1}=-\kappa q^{4}\tilde{u}_{1}-g(\tilde{u}_{1}-\tilde{u}_{2})+\tilde{\sigma}_{\text{ext}}\label{eq:Fourier_EOM_Sheet1}
\end{equation}
while Eq.~(\ref{eq:EOM_Sheetn}) yields 
\begin{equation}
-\rho\omega^{2}\tilde{u}_{n}=-\kappa q^{4}\tilde{u}_{n}-g(\tilde{u}_{n}-\tilde{u}_{n-1})-g(\tilde{u}_{n}-\tilde{u}_{n+1})\ .\label{eq:Fourier_EOM_Sheetn}
\end{equation}
We note here that when h-BN is strained, its flexural phonon dispersion
becomes linear in the long wavelength limit as a result of stiffening
of the flexural modes~\citep{SLi:SciRep17_Thermal} like in graphene~\citep{NBonini:NL12_Acoustic}
and a term proportional to $q^{2}$ should be added to the RHS of
Eqs.~(\ref{eq:Fourier_EOM_Sheet1}) and (\ref{eq:Fourier_EOM_Sheetn}).
However, this is unlikely to affect the TBC between the 2DLM and h-BN
significantly because, as we shall see later, the low-frequency portion
of the transmission spectrum is not significant.

Proceeding further with our analysis, we combine Eqs.~(\ref{eq:Fourier_EOM_Sheet1})
and (\ref{eq:Fourier_EOM_Sheetn}) to obtain the matrix equation 
\begin{align}
-z(\boldsymbol{q},\omega)\tilde{\boldsymbol{u}}(\boldsymbol{q},\omega)=-\boldsymbol{H}\tilde{\boldsymbol{u}}(\boldsymbol{q},\omega)+\tilde{\boldsymbol{\sigma}}(\boldsymbol{q},\omega)\ ,\label{eq:Fourier_EOM_MatrixForm}
\end{align}
where $z(\boldsymbol{q},\omega)=\rho\omega^{2}-\kappa q^{4}$ represents
the inverse of the response function of a decoupled h-BN layer, $\tilde{\boldsymbol{u}}(\boldsymbol{q},\omega)=(\begin{array}{cccc}
\tilde{u}_{1} & \tilde{u}_{2} & \tilde{u}_{3} & \ldots\end{array})^{T}$, $\tilde{\boldsymbol{\sigma}}(\boldsymbol{q},\omega)=(\begin{array}{cccc}
\tilde{\sigma}_{\text{ext}} & 0 & 0 & \ldots\end{array})^{T}$ , and 
\[
\boldsymbol{H}=\left(\begin{array}{cccc}
g & -g & 0 & \ldots\\
-g & 2g & -g & \ldots\\
0 & -g & 2g & \ddots\\
\vdots & \vdots & \ddots & \ddots
\end{array}\right)
\]
is a matrix describing the interlayer coupling. We can rewrite Eq.~(\ref{eq:Fourier_EOM_MatrixForm})
as 
\begin{equation}
\tilde{\boldsymbol{u}}(\boldsymbol{q},\omega)=-\boldsymbol{D}(\boldsymbol{q},\omega)\tilde{\boldsymbol{\sigma}}(\boldsymbol{q},\omega)\label{eq:SurfaceResponseRelationship}
\end{equation}
where 
\begin{equation}
\boldsymbol{D}(\boldsymbol{q},\omega)=\left(\begin{array}{cccc}
D_{11}(\boldsymbol{q},\omega) & D_{12}(\boldsymbol{q},\omega) & D_{31}(\boldsymbol{q},\omega) & \ldots\\
D_{21}(\boldsymbol{q},\omega) & D_{22}(\boldsymbol{q},\omega) & D_{32}(\boldsymbol{q},\omega) & \ldots\\
D_{31}(\boldsymbol{q},\omega) & D_{32}(\boldsymbol{q},\omega) & D_{33}(\boldsymbol{q},\omega) & \ddots\\
\vdots & \vdots & \ddots & \ddots
\end{array}\right)\label{eq:Stack_Response}
\end{equation}
is the response function of the semi-infinite stack, i.e., $\boldsymbol{D}(\boldsymbol{q},\omega)=[z(\boldsymbol{q},\omega)\boldsymbol{I}-\boldsymbol{H}]^{-1}$,
and $\boldsymbol{I}$ is the identity matrix. Most of the matrix elements
of $\boldsymbol{D}(\boldsymbol{q},\omega)$ are not relevant for our
problem since we only need $D_{11}(\boldsymbol{q},\omega)$, the first
diagonal matrix element in Eq.~(\ref{eq:Stack_Response}), to determine
how $\tilde{u}_{1}(\boldsymbol{q},\omega)$ responds to $\tilde{\sigma}_{\text{ext}}(\boldsymbol{q},\omega)$
because Eq.~(\ref{eq:SurfaceResponseRelationship}) implies that
$\tilde{u}_{1}(\boldsymbol{q},\omega)=-D_{11}(\boldsymbol{q},\omega)\tilde{\sigma}_{\text{ext}}(\boldsymbol{q},\omega)$.
We note that because the matrices and column vectors in Eq.~(\ref{eq:Fourier_EOM_MatrixForm})
are infinitely large, it is impossible to determine $D_{11}(\boldsymbol{q},\omega)$
from direct matrix inversion. Instead, an alternative approach is
needed.

\subsubsection{Analytical expression for $D_{11}(\boldsymbol{q},\omega)$}

As shown earlier in Ref.~\citep{ZYOng:PRB16_Theory}, $D_{11}(\boldsymbol{q},\omega)$
must satisfy the relationship
\begin{equation}
D_{11}(\boldsymbol{q},\omega)=[z(\boldsymbol{q},\omega)-P(\boldsymbol{q},\omega)]^{-1}\label{eq:D11_with_selfenergy}
\end{equation}
where $P(\boldsymbol{q},\omega)=g[1-gD_{22}^{(0)}(\boldsymbol{q},\omega)]^{-1}$
is the `self-energy' term representing the effect of the subsurface
layers ($n\geq2$) coupling to the top layer ($n=1$), and $D_{22}^{(0)}$
is the second diagonal matrix element in $\boldsymbol{D}_{0}(\boldsymbol{q},\omega)=[z(\boldsymbol{q},\omega)\boldsymbol{I}-\boldsymbol{H}_{0}]^{-1}$
for 
\[
\boldsymbol{H}_{0}=\left(\begin{array}{cccc}
0 & 0 & 0 & \ldots\\
0 & g & -g & \ldots\\
0 & -g & 2g & \ddots\\
\vdots & \vdots & \ddots & \ddots
\end{array}\right)\ .
\]
Physically, $D_{22}^{(0)}$ represents the response of the semi-infinite
stack ($n=2,3,\ldots$) in layer 2 when layer $1$ is decoupled from
the layer 2. This decoupling effectively turns layer 2 into the surface
layer of a semi-infinite stack and we thus deduce that $D_{22}^{(0)}(\boldsymbol{q},\omega)=D_{11}(\boldsymbol{q},\omega)$.
Hence, we obtain for Eq.~(\ref{eq:D11_with_selfenergy}) the expression
\begin{equation}
D_{11}(\boldsymbol{q},\omega)=\left[z(\boldsymbol{q},\omega)-\frac{g}{1-gD_{11}(\boldsymbol{q},\omega)}\right]^{-1}\ ,\label{eq:D11_expression}
\end{equation}
which we rewrite as an equation quadratic in $D_{11}$, i.e., 
\begin{equation}
D_{11}(\boldsymbol{q},\omega)[1-gD_{11}(\boldsymbol{q},\omega)]=z(\boldsymbol{q},\omega)^{-1}\ .\label{eq:D11_asymptotic_analysis}
\end{equation}
The solution to Eq.~(\ref{eq:D11_asymptotic_analysis}) is 
\begin{equation}
D_{11}^{\pm}(\boldsymbol{q},\omega)=\frac{2}{z(\boldsymbol{q},\omega)\pm\sqrt{z(\boldsymbol{q},\omega)^{2}-4z(\boldsymbol{q},\omega)g}}\ .\label{eq:D11_plus_minus}
\end{equation}

However, $D_{11}^{+}(\boldsymbol{q},\omega)$ and $D_{11}^{-}(\boldsymbol{q},\omega)$
are not the solutions to Eq.~(\ref{eq:D11_expression}) for $z\in\mathbb{R}$
because they do not satisfy the constraints on their asymptotic behavior
imposed by Eq.~(\ref{eq:D11_asymptotic_analysis}). We also observe
that $D_{11}^{\pm}(\boldsymbol{q},\omega)$ has a singularity at $z=0$,
where the denominator in Eq.~(\ref{eq:D11_plus_minus}) is zero,
corresponding to the surface states of the semi-infinite stack model.
Thus, we have to treat $D_{11}(\boldsymbol{q},\omega)$ differently
for $z>0$ and $z<0$. For $z\rightarrow\pm\infty$, Eq.~(\ref{eq:D11_asymptotic_analysis})
has to scale asymptotically as $\lim_{z\rightarrow\pm\infty}D_{11}(\boldsymbol{q},\omega)=z(\boldsymbol{q},\omega)^{-1}$,
a condition which can only be satisfied if we choose $D_{11}(\boldsymbol{q},\omega)=D_{11}^{-}(\boldsymbol{q},\omega)$
for $z<0$ and $D_{11}(\boldsymbol{q},\omega)=D_{11}^{+}(\boldsymbol{q},\omega)$
for $z>0$. Therefore, the surface response function is 
\begin{equation}
D_{\text{sub}}(\boldsymbol{q},\omega)=\Theta(z)D_{11}^{+}(\boldsymbol{q},\omega)+\Theta(-z)D_{11}^{-}(\boldsymbol{q},\omega)\ ,\label{eq:D11_final_solution}
\end{equation}
where $\Theta(\ldots)$ is the Heaviside function.

If $0<z(\boldsymbol{q},\omega)\leq4g$, then $D_{\text{sub}}(\boldsymbol{q},\omega)$
has an imaginary component and can be written as 
\begin{equation}
D_{\text{sub}}(\boldsymbol{q},\omega)=\frac{2}{z(\boldsymbol{q},\omega)+i\sqrt{4z(\boldsymbol{q},\omega)g-z(\boldsymbol{q},\omega)^{2}}}\ ,\label{eq:D11_with_imaginary}
\end{equation}
where $z(\boldsymbol{q},\omega)=\rho\omega^{2}-\kappa q^{4}$, to
give us \begin{subequations}
\begin{equation}
\text{Re}D_{\text{sub}}(\boldsymbol{q},\omega)=\frac{1}{2g}\label{eq:Real_D11}
\end{equation}
 and 
\begin{equation}
\text{Im}D_{\text{sub}}(\boldsymbol{q},\omega)=-\frac{\sqrt{4z(\boldsymbol{q},\omega)g-z(\boldsymbol{q},\omega)^{2}}}{2z(\boldsymbol{q},\omega)g}\ .\label{eq:Imaginary_D11}
\end{equation}
\end{subequations}Outside of the range $0<z(\boldsymbol{q},\omega)\leq4g$,
the expression for $D_{\text{sub}}(\boldsymbol{q},\omega)$ in Eq.~(\ref{eq:D11_final_solution})
has no imaginary component for $z(\boldsymbol{q},\omega)<0$ or $z(\boldsymbol{q},\omega)>4g$.
The finiteness of $\text{Im}D_{\text{sub}}(\boldsymbol{q},\omega)$
when $0<z(\boldsymbol{q},\omega)\leq4g$ is a result of the existence
of bulk flexural modes which satisfy the dispersion relation 
\begin{equation}
\omega(q,k)=\sqrt{\frac{\kappa q^{4}}{\rho}+\frac{4g}{\rho}\sin^{2}\left(\frac{ka}{2}\right)\ ,}\label{eq:BulkFlexuralDispersion}
\end{equation}
where $k$ is the wave vector associated with periodicity in the cross-plane
($z$) direction. Equation~(\ref{eq:BulkFlexuralDispersion}), which
can be derived from the phonon dispersion of the one-dimensional monoatomic
lattice chain~\citep{GChen:Book05_Nanoscale}, also implies that
$\omega\leq\sqrt{\frac{1}{\rho}(\kappa q_{c}^{4}+4g)}$.

Physically, the nonzero $\text{Im}D_{\text{sub}}(\boldsymbol{q},\omega)$
implies that the surface displacement from the applied external force
$\tilde{\sigma}_{\text{ext}}(\boldsymbol{q},\omega)$ dissipates into
the h-BN substrate as bulk flexural waves. This is the mechanism for
interfacial heat transfer between the 2DLM and h-BN since $\tilde{\sigma}_{\text{ext}}(\boldsymbol{q},\omega)$
is generated from the harmonic forces at the interface and the energy
from this interaction is dissipated into the substrate bulk.

\subsubsection{Numerical verification of $D_{\text{sub}}(\boldsymbol{q},\omega)$
formula}

To verify that Eq.~(\ref{eq:D11_final_solution}) is correct numerically,
we make use of Eq.~(\ref{eq:D11_with_selfenergy}) to define the
$n$-th order approximation to $D_{\text{sub}}(\boldsymbol{q},\omega)$
as
\begin{equation}
D_{\text{sub}}^{(n)}(\boldsymbol{q},\omega)=\left[z(\boldsymbol{q},\omega)-\frac{g}{1-gD_{\text{sub}}^{(n-1)}(\boldsymbol{q},\omega)}\right]^{-1}\label{eq:D11_nth_order}
\end{equation}
for $n\geq1$ and $D_{\text{sub}}^{(0)}(\boldsymbol{q},\omega)=z(\boldsymbol{q},\omega)^{-1}$.
Physically, $D_{\text{sub}}^{(n)}(\boldsymbol{q},\omega)$ can be
interpreted as the surface response function for a finite stack with
$n+1$ layers~\citep{ZYOng:PRB17_Thickness}. We expect Eq.~(\ref{eq:D11_nth_order})
to converge to Eq.~(\ref{eq:D11_final_solution}) as we iterate it
over $n$, i.e., $\lim_{n\rightarrow\infty}D_{\text{sub}}^{(n)}=D_{\text{sub}}$.

\begin{figure}
\begin{centering}
\includegraphics[width=7.5cm]{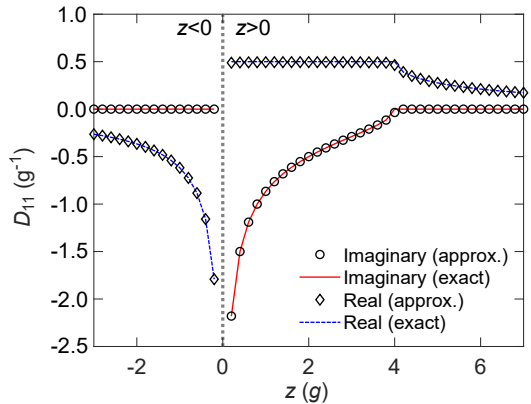}
\par\end{centering}
\caption{Plot of $\text{Re}D_{\text{sub}}$ (dashed blue line), $\text{Im}D_{\text{sub}}$
(solid red line), $\text{Re}D_{\text{sub}}^{(n)}$ (diamond symbols),
and $\text{Im}D_{\text{sub}}^{(n)}$ (circle symbols) from Eqs.~(\ref{eq:D11_final_solution})
and (\ref{eq:D11_nth_order}), respectively, for $-3g\protect\leq z\protect\leq7g$
at $n=2000$. The $z=0$ point is excluded.}

\label{fig:D11_comparisons}
\end{figure}

In our numerical tests to check if $\lim_{n\rightarrow\infty}D_{\text{sub}}^{(n)}=D_{\text{sub}}$,
we treat $D_{\text{sub}}$ and $D_{\text{sub}}^{(n)}$ as functions
of $z$ which can be either positive or negative. We compute $D_{\text{sub}}$
and $D_{\text{sub}}^{(n)}$ for $-3g\leq z\leq7g$, with the $z=0$
point excluded. To ensure convergence in the computation of $D_{\text{sub}}^{(n)}$
when $0<z\leq4g$, we make the substitution $z\rightarrow z+i\eta(z)$,
where $\eta(z)=0.01z$, in Eq.~(\ref{eq:D11_nth_order}). We find
that convergence is easily achieved at relatively small values of
$n$ ($n\sim20$) for $z<0$ and $z>4g$ where $\text{Im}D_{\text{sub}}(\boldsymbol{q},\omega)=0$.
However, in the $0<z\leq4g$ range where $\text{Im}D_{\text{sub}}\neq0$,
much higher values of $n$($n\apprge1500$) are needed for convergence
especially at smaller $z$ values. Figure~\ref{fig:D11_comparisons}
shows $D_{\text{sub}}^{(n)}$ at $n=2000$ and $D_{\text{sub}}$ for
comparison. We observe excellent agreement between the approximate
and analytical results from Eqs.~(\ref{eq:D11_nth_order}) and (\ref{eq:D11_final_solution}),
respectively, verifying that formula for $D_{\text{sub}}(\boldsymbol{q},\omega)$
in Eq.~(\ref{eq:D11_final_solution}) is correct.

\subsection{Elasticity-based models of multilayer 2D layered materials and amorphous
SiO$_{2}$}

In order to apply Eq.~(\ref{eq:ModalTransmission}) for an $N$-layer
2DLM, we also need the expression for $D_{\text{2D}}(\boldsymbol{q},\omega)$
for different values of $N$. Although the formulas are given in Ref.~\citep{ZYOng:PRB17_Thickness},
we provide them here for the sake of completeness. The expression
for the $D_{\text{2D}}(\boldsymbol{q},\omega)$ of a 1-layer 2DLM
is~\citep{ZYOng:PRB16_Theory} 
\begin{equation}
D_{\text{2D},1}(\boldsymbol{q},\omega)=\frac{1}{\rho\omega^{2}+i\rho\gamma(\omega,T)\omega-\kappa q^{4}}\label{eq:FlexuralResponse_1Layer}
\end{equation}
where $\rho$ is the areal mass density, $\gamma(\omega)$ is the
frequency-dependent damping function, and $\kappa$ is the bending
rigidity, for a single-layer 2DLM. At temperature $T$, we can approximate
the damping function~\citep{ZYOng:PRB16_Theory} as $\gamma(\omega,T)=\frac{\omega T}{\alpha T_{\text{RT}}}$,
where $T_{\text{RT}}$ is the room temperature and $\alpha$ is the
ratio of the flexural mode frequency to its inverse lifetime at $T=T_{\text{RT}}$.
Phenomenologically, the frequency-dependent damping function $\gamma(\omega,T)$
is proportional to $\omega$~\citep{LLindsay:PRB14_Phonon} and represents
the coupling and exchange of energy between the flexural phonons and
other intrinsic degrees of freedom in the 2DLM. In the language of
many-body physics, $\gamma(\omega,T)$ corresponds to the self-energy
from the interaction (e.g. anharmonic phonon-phonon and electron-phonon
coupling) of the flexural phonons with other intrinsic degrees of
freedom and plays an analogous role to the B\"uttiker probe self-energy
terms used in Ref.~\citep{SSadasivam:PRB17_Thermal}. The temperature
dependence of $\gamma(\omega,T)$ means that $D_{\text{2D}}(\boldsymbol{q},\omega)$
and hence the total transmission function $\xi(\omega)$ in Eq.~(\ref{eq:TotalTransmission})
are temperature-dependent. In 2DLM's with defects, $\gamma(\omega,T)$
can also include the effects of defect scattering~\citep{CAPolanco:PRB18_AbInitio}
which increases the damping rate and leads to higher dissipation rates.
For an $N$-layer 2DLM ($N>1$), the corresponding expression for
the flexural response at the 2DLM-substrate interface is~\citep{ZYOng:PRB17_Thickness}
\begin{equation}
D_{\text{2D},N}(\boldsymbol{q},\omega)=\sum_{n=1}^{N}\frac{f_{n}}{D_{\text{2D},1}(\boldsymbol{q},\omega)^{-1}-z_{n}}\label{eq:FlexuralResponse_NLayer}
\end{equation}
where $z_{n}=4g\sin^{2}[\frac{(n-1)\pi}{2N}]$, $g$ is the interlayer
spring constant, $f_{1}=\frac{1}{N}$, and $f_{n}=\frac{1}{N}[1+\cos\frac{(n-1)\pi}{N}]$
for $1<n\leq N$. We can estimate $g$ from the $\Gamma$-point interlayer
breathing modes and obtain $g=10.95\times10^{19}$ N/m$^{3}$ and
$g=8.60\times10^{19}$ N/m$^{3}$ for graphene~\citep{ZYOng:PRB17_Thickness}
and MoS$_{2}$~\citep{YZhao:NL13_Interlayer,LLiang:ACSNano17_LowFrequency},
respectively. In the $N\rightarrow\infty$ limit, we recover the expression
in Eq.~(\ref{eq:D11_final_solution}), i.e., 
\begin{align}
\lim_{\gamma\rightarrow0}D_{\text{2D},\infty}(\boldsymbol{q},\omega)=\frac{\Theta(\zeta)}{\zeta+\sqrt{\zeta^{2}-4\zeta g}}+\frac{\Theta(-\zeta)}{\zeta-\sqrt{\zeta^{2}-4\zeta g}}\label{eq:FlexuralResponse_InfLayer}
\end{align}
where $\zeta=\lim_{\gamma\rightarrow0}D_{\text{2D},1}(\boldsymbol{q},\omega)^{-1}$.
To model the interface between an $N=\infty$ 2DLM and its substrate,
we set $\gamma(\omega,T)=0$ so that $D_{\text{2D}}(\boldsymbol{q},\omega)=\lim_{\gamma\rightarrow0}D_{\text{2D},\infty}(\boldsymbol{q},\omega)$
like in Eq.~(\ref{eq:FlexuralResponse_InfLayer}). The dissipative
term $\gamma(\omega,T)$ is eliminated in Eq.~(\ref{eq:FlexuralResponse_InfLayer})
because the semi-infinite structure of the $N=\infty$ 2DLM ensures
that $D_{\text{2D}}(\boldsymbol{q},\omega)$ is dissipative.

To model the surface Green's function for the h-BN substrate, we use
the parameters $\rho=7.67\times10^{-7}$ kg/m$^{2}$, $\kappa=0.86$
eV~\citep{SSingh:PRB13_Thermomechanical} and $g=9.83\times10^{19}$
N/m$^{3}$~\citep{LLiang:ACSNano17_LowFrequency} in Eq.~(\ref{eq:D11_final_solution}).
We assume the maximum wave vector for h-BN to be $q_{\text{max}}=2.7\times10^{10}$
m$^{-1}$ and use Eq.~(\ref{eq:BulkFlexuralDispersion}) to set $\omega_{\text{max}}=\sqrt{\frac{1}{\rho}(\kappa q_{\text{cut}}^{4}+4g)}=204$
meV as the cutoff frequency above which $D_{\text{sub}}(\boldsymbol{q},\omega)=0$.
To model the surface Green's function for the a-SiO$_{2}$ substrate,
we approximate it as an isotropic elastic solid for which its surface
Green's function is~\citep{BNJPersson:JPCM11_Phononic,ZYOng:PRB16_Theory}
\begin{equation}
D_{\text{sub}}(\boldsymbol{q},\omega)=\frac{i\omega^{2}}{\rho_{\text{sub}}c_{T}^{4}}\frac{p_{L}(q,\omega)}{S(q,\omega)}\Theta(\omega_{D}-\omega)\label{eq:SurfaceResponse_SiO2}
\end{equation}
where $S(q,\omega)=[(\omega/c_{T})^{2}-2q^{2}]^{2}+4q^{2}p_{T}(q,\omega)p_{L}(q,\omega)$,
$p_{L}(q,\omega)=\lim_{\eta\rightarrow0^{+}}\sqrt{(\omega/c_{L})^{2}-q^{2}+i\eta}$,
and $p_{T}(q,\omega)=\lim_{\eta\rightarrow0^{+}}\sqrt{(\omega/c_{T})^{2}-q^{2}+i\eta}$,
with $\rho_{\text{sub}}$ and $c_{L}$ ($c_{T}$) denoting the voluminal
mass density and the longitudinal (transverse) speed of sound in a-SiO$_{2}$,
respectively. We set $\rho_{\text{sub}}=2200$ kg/m$^{3}$, $c_{L}=5953$
m/s and $c_{T}=2200$ m/s following Ref.~\citep{ZYOng:PRB16_Theory}.
The longitudinal Debye frequency is $\omega_{D}=(6\pi^{2}N_{\text{sub}}c_{L}^{3})^{1/3}=62$
meV where $N_{\text{sub}}=6.62\times10^{28}$ m$^{-3}$ is the number
density for a-SiO$_{2}$~\citep{KGoodson:JHT94_Prediction}. To model
the harmonic interaction at the interface of the 2DLM and h-BN, we
estimate $K$ from the arithmetic mean of the interlayer spring constant
$g$ for the 2DLM and h-BN.

\begin{table}
\begin{centering}
\begin{tabular}{|c|c|c|}
\hline 
 & Graphene & MoS$_{2}$\tabularnewline
\hline 
$\rho$ (kg/m$^{2}$) & $7.63\times10^{-7}$ & $3.05\times10^{-6}$\tabularnewline
\hline 
$\kappa$ (eV) & $1.1$ & $9.61$\tabularnewline
\hline 
$g$ ($10^{19}$ N/m$^{3}$) & $10.95$ & $8.60$\tabularnewline
\hline 
\multirow{2}{*}{$K$ ($10^{19}$ N/m$^{3}$)} & $12.30$ (SiO$_{2}$) & $4.94$ (SiO$_{2}$)\tabularnewline
 & $10.39$ (h-BN) & $9.21$ (h-BN)\tabularnewline
\hline 
$\alpha$ & 100 & 100\tabularnewline
\hline 
$q_{c}$ ($10^{10}$ m$^{-1}$) & $2.7$ & $2.1$\tabularnewline
\hline 
\end{tabular}
\par\end{centering}
\caption{Simulation parameters for graphene and MoS$_{2}$ taken from Refs.~\citep{ZYOng:PRB16_Theory,ZYOng:PRB17_Thickness,LLiang:ACSNano17_LowFrequency}.
The values of $K$ for the graphene-SiO$_{2}$ and MoS$_{2}$-SiO$_{2}$
interface are taken from Ref.~\citep{ZYOng:PRB16_Theory} while that
for the graphene-BN (MoS$_{2}$-BN) interface is estimated from the
arithmetic mean of the $g$ for graphene and h-BN (MoS$_{2}$ and
h-BN) with $g=9.83\times10^{19}$ N/m$^{3}$ for h-BN~\citep{LLiang:ACSNano17_LowFrequency}
.}
\label{tab:2DLM_sim_parameters}
\end{table}

\section{Results and discussion}

We compute the TBC numerically, using Eq.~(\ref{eq:LandauerTBCFormula}),
for different 2DLMs (graphene and MoS$_{2}$) and substrates (h-BN
and a-SiO$_{2}$). We use a-SiO$_{2}$ as a contrast in order to understand
the effect of the layered geometry of h-BN on cross-plane substrate-directed
heat dissipation from two common 2DLMs. In using Eq.~(\ref{eq:LandauerTBCFormula}),
we assume for the sake of simplicity that the temperature dependence
of the TBC is due primarily to the temperature-dependent changes in
the Bose-Einstein distribution $N(\omega,T)$ of the phonons and the
damping function $\gamma(\omega,T)$ in Eq.~(\ref{eq:FlexuralResponse_1Layer}).

Here, a few remarks on the effects on anharmonicity are necessary
for understanding the limitations of our TBC results. Although it
is known that anharmonicity or inelastic phonon scattering~\citep{PHopkins:JHT11}
can play a role in interfacial thermal transport~\citep{NMingo:PRB06_Anharmonic,PHopkins:JHT11,JDai:PRB20_Rigorous},
the degree to which it contributes to the TBC remains an open question.
Anharmonicity has two attendant effects on interfacial transport:
at the interface, anharmonicity enables the decomposition of higher-energy
phonons on one side of the interface into two or more lower-energy
phonons on the other side~\citep{PHopkins:JHT11} while in the bulk
of a solid, the anharmonic coupling between the phonons leads to the
bulk phonons having a finite lifetime. In this study, we do not study
the first effect as a rigorous treatment is beyond the scope and objective
of our work. Only the second effect, the finite lifetime of the phonons,
is taken into account in our results. It is included indirectly and
partially, albeit in a phenomenological fashion, through the damping
function $\gamma(\omega,T)$ in Eq.~(\ref{eq:FlexuralResponse_1Layer})
which estimates the finite phonon lifetime in the 2DLM due to inelastic
phonon scattering. Indeed, it is noted in Ref.~\citep{ZYOng:PRB16_Theory}
that for a 2DLM with a \emph{finite} number of layers, the inclusion
of a non-zero damping function is needed in order for the thermal
resistance of the boundary to be finite, i.e., the flexural modes
in the 2DLM cannot dissipate heat to the substrate unless the individual
layers have some form of damping.

On the other hand, we make no attempt to model the effects of anharmonicity
in the h-BN substrate where anharmonic scattering of the bulk phonons
can occur~\citep{LLindsay:PRB11_Enhanced}. This is also because
the flexural motion of the surface of the h-BN substrate is naturally
damped via its harmonic coupling to the subsurface layers in the h-BN,
an effect represented by the `self-energy' term $P(\boldsymbol{q},\omega)$
in Eq.~(\ref{eq:D11_with_selfenergy}). Although it is possible to
include a phenomenological damping term in our derivation of $D_{\text{sub}}$
in Eq.~(\ref{eq:D11_final_solution}), its inclusion would result
in the intractability of the derivation and analysis of $D_{\text{sub}}$.
Moreover, the necessity of including anharmonicity in h-BN remains
unclear to us since the theoretical interpretation and analysis of
existing experimental data~\citep{JGaskins:NL18_Thermal,ZCheng:CommunPhys20_Thermal}
provide no guidance on its inclusion. While discrepancies between
purely elastic calculations and experimental measurements of the TBC,
such as those for the ZnO/GaN interface~\citep{JGaskins:NL18_Thermal},
suggest that anharmonic effects may play a significant role in the
TBC at high temperatures, other results~\citep{ZCheng:CommunPhys20_Thermal}
indicate that the contribution to the TBC from inelastic processes
can be insignificant even at high temperatures. These differences
imply that the importance of interfacial thermal transport may vary
with the specific material interface even if anharmonic scattering
is important for bulk heat conduction.

\subsection{Temperature dependence of the TBC}

\begin{figure}
\begin{centering}
\includegraphics[scale=0.38]{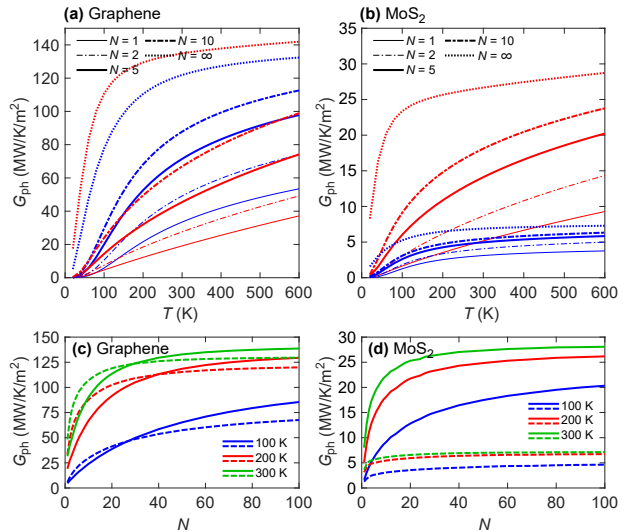}
\par\end{centering}
\caption{Plot of the TBC $G_{\text{ph}}$ for single and multi-layer (a) graphene
and (b) MoS$_{2}$ with a-SiO$_{2}$ (blue lines) or h-BN (red lines)
as the substrate over the temperature range of $T=20$ to $600$ K.
The TBC is shown for $N=1$, $2$, $5$, $10$, and $\infty$. We
also plot the layer-dependent TBC for (c) graphene and (d) MoS$_{2}$
with a-SiO$_{2}$ (dashed lines) or h-BN (solid lines) as the substrate
at $T=100$, $200$ and $300$ K.}
\label{fig:TempDependTBC}
\end{figure}

We compare in Fig.~\ref{fig:TempDependTBC} the temperature dependence
of the TBC $G_{\text{ph}}$ for the 2DLM-substrate interface with
single ($N=1$) and few-layer ($N=2,5,10$) graphene or MoS$_{2}$
as the 2DLM and a-SiO$_{2}$ or h-BN as the substrate over the temperature
range of $T=20$ to $600$ K. In addition, we also plot the $G_{\text{ph}}$
results for the $N=\infty$ case where the 2DLM is the bulk form of
graphene and MoS$_{2}$, corresponding respectively to graphite and
molybdenite. In Fig.~\ref{fig:TempDependTBC}(a), we find that the
graphene-substrate TBC increases with $T$ as well as the number of
layers in the 2DLM $N$ like in Ref.~\citep{ZYOng:PRB17_Thickness},
reaching $G_{\text{ph}}=135$ MW/K/m$^{2}$ (graphite-BN) and $G_{\text{ph}}=122$
MW/K/m$^{2}$ (graphite-SiO$_{2}$) in the $N=\infty$ limit at 300
K. This pronounced $N$-dependence of the TBC for the graphene-BN
interface suggests that the thickness-dependent TBC is a general phenomenon
that extends to a wider class of substrates beyond elastically isotropic
ones such as SiO$_{2}$~\citep{ZYOng:PRB17_Thickness}. For $N\leq10$,
the graphene-SiO$_{2}$ TBC is also higher than the graphene-BN TBC,
with $G_{\text{ph}}=33.5$ MW/K/m$^{2}$ and $G_{\text{ph}}=19.1$
MW/K/m$^{2}$ for the former and latter respectively at $N=1$. This
suggests that SiO$_{2}$ may be more efficient as a vertical heat
sink for few-layer graphene although the significantly lower thermal
conductivity of SiO$_{2}$ ($\sim1.3$ to $1.5$ W/m/K~\citep{DCahill:ARPC88_Lattice,SLee:JAP97_Heat,TYamane:JAP02_Measurement,JMLarkin:PRB14_Amorphous})
may make it a suboptimal choice for overall heat dissipation. 

In Fig.~\ref{fig:TempDependTBC}(b), we find that the MoS$_{2}$-substrate
TBC also increases with $T$ and $N$. The MoS$_{2}$-BN TBC is however
much larger than the MoS$_{2}$-SiO$_{2}$ TBC for all temperatures
unlike in graphene. At $T=300$ K, we obtain $G_{\text{ph}}=5.2$
MW/K/m$^{2}$ (MoS$_{2}$-BN) and $G_{\text{ph}}=3.1$ MW/K/m$^{2}$
(MoS$_{2}$-SiO$_{2}$) for $N=1$, and $G_{\text{ph}}=26.7$ MW/K/m$^{2}$
(MoS$_{2}$-BN) and $G_{\text{ph}}=6.9$ MW/K/m$^{2}$ (MoS$_{2}$-SiO$_{2}$)
in the $N=\infty$ limit. Our result of $G_{\text{ph}}=26.7$ MW/K/m$^{2}$
for $N=\infty$ and $T=300$ K is comparable to the result of $20$
MW/K/m$^{2}$ obtained in Ref.~\citep{YLiu:SciRep17_Thermal} using
the Atomistic Green's Function method, which we discuss in Sec.~\ref{Subsec:Comparison_with_AGF}.
At $T=300$ K and $N=1$, the ratio of the graphene-BN TBC to the
MoS$_{2}$-BN TBC is $16.1/5.2\sim3.1$, in good agreement with the
$\sim3.1$ ratio obtained in thermometric measurements using Raman
spectroscopy~\citep{YLiu:SciRep17_Thermal}. The much larger TBC
of the MoS$_{2}$-BN interface implies that h-BN can be a more effective
substrate material than SiO$_{2}$ for heat dissipation from MoS$_{2}$.
A summary comparing the TBC results is shown in Table~\ref{Table:TBCResultsSummary}.
In addition, we also show the layer dependence of the TBC with the
a-SiO$_{2}$ and h-BN substrates in Figs.~\ref{fig:TempDependTBC}(c)
and (d) for graphene and MoS$_{2}$, respectively, at $T=100$, $200$
and $300$ K.

\begin{table*}
\begin{tabular}{|c|c|c|c|}
\hline 
Interface & \multicolumn{3}{c|}{TBC (MW/K/m$^{2}$)}\tabularnewline
\hline 
 & Stack model & AGF & Experiment\tabularnewline
\hline 
\hline 
Graphene-BN ($N=1$) & 19.1 & \textendash{} & 52.2~\citep{YLiu:SciRep17_Thermal}, 7.6~\citep{CChen:APL14_Thermal},10~\citep{DKim:2DM18_Energy}\tabularnewline
\hline 
MoS$_{2}$-BN ($N=1$) & 5.2 & \textendash{} & 17.0~\citep{YLiu:SciRep17_Thermal}\tabularnewline
\hline 
Graphene-BN ($N=\infty$) & 135 & 108~\citep{YLiu:SciRep17_Thermal} & \textendash{}\tabularnewline
\hline 
MoS$_{2}$-BN ($N=\infty$) & 26.7 & 20~\citep{YLiu:SciRep17_Thermal} & \textendash{}\tabularnewline
\hline 
\end{tabular}

\caption{Comparison of TBC results at $T=300$ K obtained for the various $N$-layer
2DLM-BN interfaces using theoretical and experimental techniques.}

\label{Table:TBCResultsSummary}
\end{table*}

\begin{figure}
\begin{centering}
\includegraphics[scale=0.38]{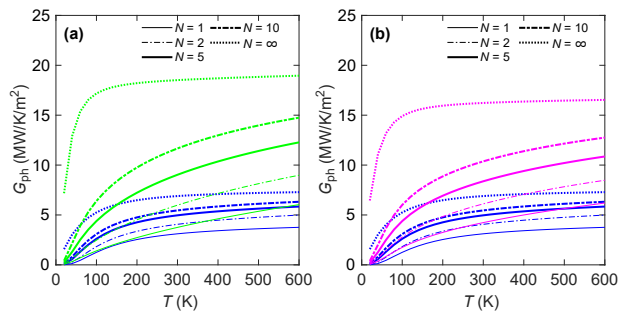}
\par\end{centering}
\caption{(a) Plot of the TBC $G_{\text{ph}}$ for single and multi-layer MoS$_{2}$
with a-SiO$_{2}$ (blue lines) or h-BN (green lines) as the substrate
over the temperature range of $T=20$ to $600$ K, assuming $K=4.94\times10^{19}$
N/m$^{3}$ for both the MoS$_{2}$-BN and the MoS$_{2}$-SiO$_{2}$
interfaces. The TBC is shown for $N=1$, $2$, $5$, $10$, and $\infty$.
(b) Same plot of $G_{\text{ph}}$ but with $g=2.24\times10^{20}$
N/m$^{3}$ for the interlayer areal spring constant in h-BN so that
its acoustic impedance in the normal direction is identical to that
of a-SiO$_{2}$. The MoS$_{2}$-BN results are indicated with magenta
lines. We assume that $K=4.94\times10^{19}$ N/m$^{3}$ for both the
MoS$_{2}$-BN and the MoS$_{2}$-SiO$_{2}$ interfaces}
\label{fig:TempDependTBCMoS2Comparison}
\end{figure}

Could the considerably larger $G_{\text{ph}}$ for the MoS$_{2}$-BN
interface relative to the MoS$_{2}$-SiO$_{2}$ interface be entirely
due to its significantly larger areal spring constant ($K=9.21\times10^{19}$
N/m$^{3}$ for MoS$_{2}$-BN vs. $K=4.94\times10^{19}$ N/m$^{3}$
for MoS$_{2}$-SiO$_{2}$)? We rescale the $K$ for the MoS$_{2}$-BN
interface so that it is identical to that for the MoS$_{2}$-SiO$_{2}$
interface (i.e., $K=4.94\times10^{19}$ N/m$^{3}$ for the MoS$_{2}$-BN
interface) and repeat the TBC calculations in Fig.~\ref{fig:TempDependTBC}(b),
with the results shown in Fig.~\ref{fig:TempDependTBCMoS2Comparison}(a).
Although the TBC results for the MoS$_{2}$-BN interface have decreased
significantly as a result of the reduced $K$, we find that $G_{\text{ph}}$
for the MoS$_{2}$-BN interface is still significantly higher than
$G_{\text{ph}}$ for the MoS$_{2}$-SiO$_{2}$ interface. Like in
Fig.~\ref{fig:TempDependTBC}(b), we also observe the stronger dependence
on $N$ with the MoS$_{2}$-BN interface compared to the MoS$_{2}$-SiO$_{2}$
interface. 

To sharpen our analysis of how the anisotropic layered structure in
h-BN affects the TBC, we adjust $g$, the interlayer areal spring
constant in h-BN, so that the acoustic impedance $z_{0}$ in the direction
normal to the interface is identical for h-BN and a-SiO$_{2}$. Before
the adjustment, we have $z_{0}=\rho_{\text{sub}}c_{L}=1.31\times10^{7}$
Pa$\cdot$s/m in a-SiO$_{2}$ which is higher than the $z_{0}=\sqrt{g\rho}=8.69\times10^{6}$
Pa$\cdot$s/m in h-BN. We increase the $g$ in h-BN from $9.83\times10^{19}$
N/m$^{3}$ to $2.24\times10^{20}$ N/m$^{3}$ so that its new rescaled
$z_{0}$ is equal to that of a-SiO$_{2}$. Physically, this amounts
to stronger coupling between the h-BN layers and a higher group velocity
in the normal direction. Assuming $K=4.94\times10^{19}$ N/m$^{3}$
for both the MoS$_{2}$-BN and the MoS$_{2}$-SiO$_{2}$ interfaces,
we calculate the $G_{\text{ph}}$ for $T=20$ to $600$ K and $N=1$,
$2$, $5$, $10$, and $\infty$, like in Fig.~\ref{fig:TempDependTBCMoS2Comparison}(a),
with the results shown in Fig.~\ref{fig:TempDependTBCMoS2Comparison}(b).
Compared to Fig.~\ref{fig:TempDependTBCMoS2Comparison}(a), there
is a clear decrease in $G_{\text{ph}}$, especially for $N\geq10$,
for the MoS$_{2}$-BN interface. Nonetheless, in Fig.~\ref{fig:TempDependTBCMoS2Comparison}(b),
we still observe that $G_{\text{ph}}$ for the MoS$_{2}$-BN interface
still has a greater $N$-dependence and is significantly higher than
$G_{\text{ph}}$ for the MoS$_{2}$-SiO$_{2}$ interface, further
highlighting the possible role of anisotropy in interfacial thermal
transport at the MoS$_{2}$-BN interface.

\subsection{TBC dependence on the interlayer spring constant in h-BN}

Given the sensitivity of the MoS$_{2}$-BN TBC to the numerical values
of $K$ and $g$ as seen in Fig.~\ref{fig:TempDependTBCMoS2Comparison},
we try to understand more systematically how $g$ in h-BN affects
the TBC by calculating $G_{\text{ph}}$ for $N$-layer graphene and
MoS$_{2}$ at different $g$ values. We set a natural scale for $g$
by relating it to $E,$ the in-plane elastic modulus of a h-BN monolayer,
which quantifies the relationship between its in-plane tensile stress
and axial strain. Its value, which has been measured experimentally
to be $865$ GPa~\citep{AFalin:NatCommun17_Mechanical}, can be used
to set the interlayer spring constant $g_{\text{iso}}$ which reproduces
the equivalent relationship for the cross-plane stress and strain,
i.e., $g_{\text{iso}}=E/a=2.54\times10^{21}$ N/m$^{3}$, where $a=0.34$
nm. We may regard $g_{\text{iso}}$ as the value for which cross-plane
elastic modulus is the same as the in-plane elastic modulus or the
value of $g$ in the `isotropic' limit although the stack model is
not isotropic as it only describes the out-of-plane flexural motion
of the h-BN layers.

\begin{figure}
\begin{centering}
\includegraphics[scale=0.45]{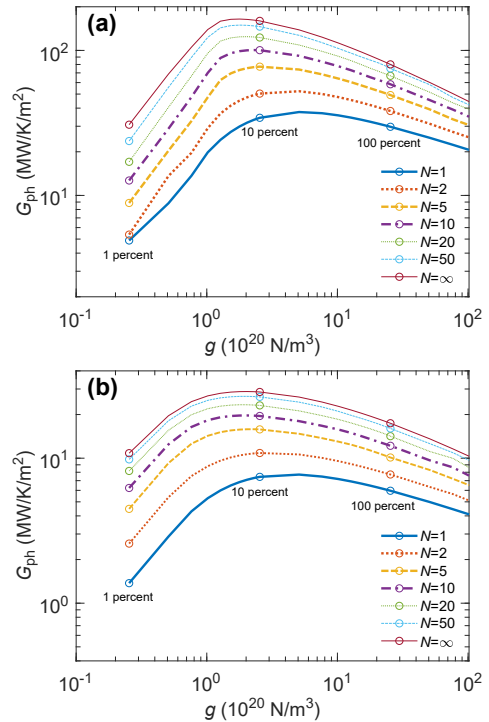}
\par\end{centering}
\caption{Plot of the TBC $G_{\text{ph}}$ at $T=300$ K for $N$-layer (a)
graphene and (b) MoS$_{2}$ with h-BN as the substrate for different
values of $g$, the interlayer spring constant in h-BN. The TBC is
shown for $N=1$, $2$, $5$, $10$, $20$, $50$ and $\infty$. We
vary the value of $g$ from 1 to 400 percent of $g_{\text{iso}}$
where $g_{\text{iso}}=2.54\times10^{21}$ N/m$^{3}$.}
\label{fig:TBC_vs_g}
\end{figure}

Figure~\ref{fig:TBC_vs_g} shows the TBC of the graphene-BN and MoS$_{2}$-BN
interfaces at $T=300$ K for different values of $g$ between 1 and
400 percent of $g_{\text{iso}}$. We observe that $G_{\text{ph}}$
decreases asymptotically when $g\gg g_{\text{iso}}$ (strong coupling
regime) and $g\ll g_{\text{iso}}$ (van der Waals regime), with $G_{\text{ph}}$
for multilayer graphene and MoS$_{2}$ peaking in the $g\sim10^{20}$
N/m$^{3}$ range, an order of magnitude smaller than $g_{\text{iso}}$.
This peak position can be estimated by comparing the cross-plane acoustic
impedance of the 2DLM, given by $z_{0}=\sqrt{\rho g_{\text{2D}}}$
where $g_{\text{2D}}$ is the interlayer spring constant in graphene
or MoS$_{2}$, with the cross-plane acoustic impedance of h-BN. Given
that $z_{0}=9.14\times10^{6}$ and $1.62\times10^{7}$ Pa$\cdot$s/m
for graphene and MoS$_{2}$, respectively, the interlayer spring constant
in h-BN has to be around $g=z_{0}^{2}/\rho_{\text{BN}}\sim10^{20}$
N/m$^{3}$, where $\rho_{\text{BN}}$ is the areal mass density of
the h-BN monolayer, in order for the acoustic impedance of h-BN to
match that of the 2DLM to maximize cross-plane transmission.

\subsection{Interfacial phonon transmission spectra}

\begin{figure}
\begin{centering}
\includegraphics[scale=0.38]{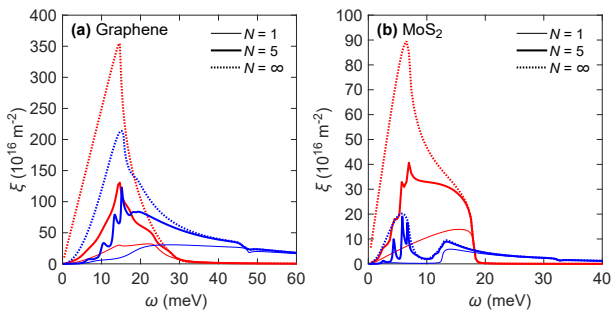}
\par\end{centering}
\caption{Plot of the interfacial transmission spectrum $\xi(\omega)$ for (a)
graphene and (b) MoS$_{2}$ with a-SiO$_{2}$ (blue lines) or h-BN
(red lines) as the substrate at $T=300$ K for $N=1$ (thin solid
lines), $5$ (thick solid lines) and $\infty$ (dotted lines).}
\label{fig:TransmitSpectra}
\end{figure}

To understand the trends in Fig.~\ref{fig:TempDependTBC}, we plot
the transmission spectra $\xi(\omega)$ from Eq.~(\ref{eq:TotalTransmission})
for $N=1,5,\infty$ layers and $T=300$ K. The increase in $G_{\text{ph}}$
with $N$ can be explained by the growth of the low-frequency transmission
spectra as $N$ increases, which permits more phonon transmission
across the 2DLM-substrate interface. The higher TBC for the graphene-SiO$_{2}$
interface relative to the graphene-BN interface in single and few-layer
graphene can be explained by its larger transmission spectra $\xi(\omega)$
at higher frequencies, as shown in Fig.~\ref{fig:TransmitSpectra}(a)
where the graphene-SiO$_{2}$ spectra are larger at higher frequencies
($\omega\apprge20$ meV) but smaller at lower frequencies ($\omega<20$
meV) than the graphene-BN transmission spectra with the relative difference
between the two spectra narrowing as $N$ increases.

Figure~\ref{fig:TransmitSpectra}(b) shows the corresponding transmission
spectra for the MoS$_{2}$-SiO$_{2}$ and MoS$_{2}$-BN interfaces
at $T=300$ K. At high frequencies ($\omega>19$ meV), $\xi(\omega)$
is greater for the MoS$_{2}$-SiO$_{2}$ interface than that for the
MoS$_{2}$-BN interface at $N=1,5$ and $\infty$. On the other hand
in the low-frequency regime ($\omega<19$ meV), $\xi(\omega)$ is
considerably greater for the MoS$_{2}$-BN interface, indicating that
h-BN is much more transparent to low-frequency phonon transmission
from MoS$_{2}$ than SiO$_{2}$ is. This explains the substantially
larger TBC of the MoS$_{2}$-BN interface compared to the MoS$_{2}$-SiO$_{2}$
interface.

\subsection{Comparison with the Atomistic Green's Function model~\label{Subsec:Comparison_with_AGF}}

\begin{figure}
\begin{centering}
\includegraphics[scale=0.38]{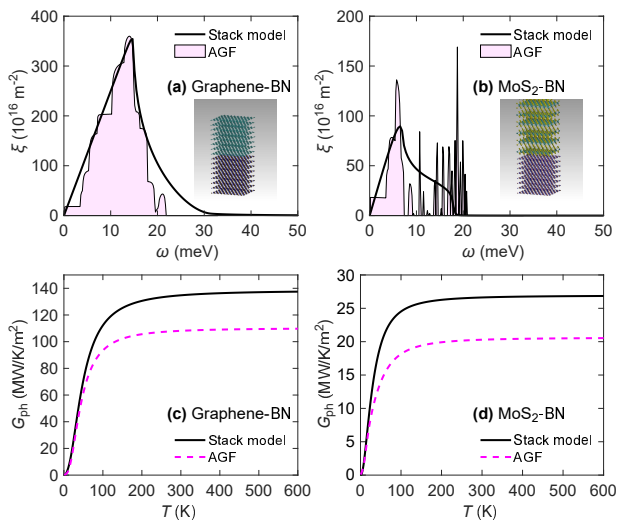}
\par\end{centering}
\caption{Comparison of the interfacial transmission spectra $\xi(\omega)$
for the (a) graphene-BN and (b) MoS$_{2}$-BN interface computed with
the stack and AGF models for $N=\infty$. The inset in each panel
shows the atomistic structure of the 2DLM-BN interface. The corresponding
temperature-dependent TBC $G_{\text{ph}}$ from $T=1$ to $600$ K
for the (c) graphene-BN and (d) MoS$_{2}$-BN interface are also shown.}
\label{fig:ComparisonAGF}
\end{figure}

In Fig.~\ref{fig:ComparisonAGF}, we compare the results for the
transmission spectra $\xi(\omega)$ and the TBC $G_{\text{ph}}$ from
our stack model with the more detailed ones from the AGF model~\citep{WZhang:NHT07_Atomistic},
which treats the interfacial thermal transport between two semi-infinite
($N=\infty$) crystal lattices, for the graphite-BN (i.e. graphene-BN
for $N=\infty$) and molybdenite-BN (i.e. MoS$_{2}$-BN for $N=\infty$)
interface. The AGF spectra are taken from Ref.~\citep{YLiu:SciRep17_Thermal}.
In our AGF simulation, the individual graphene layers are stacked
in the A-B configuration while the individual h-BN and MoS$_{2}$
layers are stacked in the A-A' configuration. The cross-sectional
area of the 2DLM interface is $2.49\times2.16$ nm$^{2}$. We note
here that the following TBC results are obtained from a purely elastic
treatment.

Figure~\ref{fig:ComparisonAGF}(a) shows the $\xi(\omega)$ for the
graphene-BN interface calculated with the stack and AGF model. Unlike
the $\xi(\omega)$ of the stack model which varies smoothly with $\omega$,
the $\xi(\omega)$ of the AGF model has discrete steps due to the
relatively small finite size of the interfacial cross section in the
AGF simulation~\citep{YLiu:SciRep17_Thermal}, which limits the number
of modes contributing to the thermal conductance. In spite of the
approximations used in our stack model, its $\xi(\omega)$ shows remarkably
good qualitative and even close quantitative agreement with the $\xi(\omega)$
of the AGF model, especially in the $\omega<20$ meV frequency range.
At higher frequencies, the transmission spectra diverge from one another
because the flexural phonon dispersion in graphene becomes non-parabolic
away from the Brillouin zone center. Nevertheless, this close agreement
between the AGF model and our stack model, which contains only flexural
modes, validates the assumptions used in the stack model and suggests
that heat conduction across the graphene-BN interface is dominated
by the coupling of flexural phonons between graphene and h-BN, with
the in-plane acoustic and optical phonons playing an insignificant
role.

Likewise in Fig.~\ref{fig:ComparisonAGF}(b), we also observe close
agreement between the MoS$_{2}$-BN transmission spectra of the the
stack and AGF models in the $\omega<7.4$ meV frequency range, indicating
the similarly dominant role of the flexural phonon coupling in interfacial
heat conduction. Above this frequency range, the difference in $\xi(\omega)$
becomes significant and we observe transmission peaks in the $\xi(\omega)$
of the AGF model, which we attribute to the possible coupling to other
acoustic and optical phonon modes because of the greater number of
degrees of freedom in the MoS$_{2}$ unit cell.

We compare the corresponding temperature-dependent TBC from Eq.~(\ref{eq:LandauerTBCFormula})
for the graphene-BN interface in Fig.~\ref{fig:ComparisonAGF}(c)
which shows relatively good agreement between the stack and the AGF
model especially at low temperatures ($T\lesssim40$ K). At $T=300$
K, we obtain $G_{\text{ph}}=135$ MW/K/m$^{2}$ and $108$ MW/K/m$^{2}$
for the stack and AGF model, respectively, i.e., the stack model overpredicts
the TBC. The higher $G_{\text{ph}}$ for the stack model is due to
its additional transmission contribution in $\xi(\omega)$ for $\omega>20$
meV. For the MoS$_{2}$-BN interface in Fig.~\ref{fig:ComparisonAGF}(d),
we have $G_{\text{ph}}=27$ MW/K/m$^{2}$ and $20$ MW/K/m$^{2}$
for the stack model and AGF method, respectively, at $T=300$ K with
the two $G_{\text{ph}}$ curves converging at low temperatures ($T\lesssim20$
K). At higher temperatures, the difference in the $G_{\text{ph}}$
is due to the larger $\xi(\omega)$ in the stack model for $\omega>7.4$
meV. Nonetheless, the relative closeness of the $G_{\text{ph}}$ between
the stack and AGF model supports and validates the description of
the elastic properties of the h-BN layers using Eqs.~(\ref{eq:EOM_Sheet1})
and (\ref{eq:EOM_Sheetn}).

\section{Summary and conclusion}

To understand the limits of heat dissipation from an $N$-layer 2DLM
(e.g. graphene or MoS$_{2}$) to its h-BN substrate, we have developed
a theory that depends on the elastic parameters of the 2DLM, h-BN
and their interface. In our theory, we use a stack model of the h-BN
substrate, in which the crystal is treated as a semi-infinite stack
of harmonically coupled thin plates to describe its surface flexural
response function $D_{\text{sub}}(\boldsymbol{q},\omega)$ to an external
force, to determine the TBC. We find that the TBC of the interface
between the 2DLM and h-BN increases with $N$, as with the case for
an isotropic solid substrate such as a-SiO$_{2}$. The increase of
the TBC with $N$ is stronger for h-BN than for a-SiO$_{2}$ especially
when the 2DLM is MoS$_{2}$. Our analysis shows that h-BN is more
transparent to low-frequency phonon transmission from the 2DLM (graphene
and MoS$_{2}$) than a-SiO$_{2}$ is. At large $N$, h-BN is a considerably
more effective substrate for heat dissipation from MoS$_{2}$ than
a-SiO$_{2}$ is because of the much greater low-frequency phonon transmission.
We also compare the predictions of the stack model in the $N=\infty$
limit for the graphene-BN and MoS$_{2}$-BN interface to those of
the AGF model. The good agreement in the low-frequency phonon transmission
spectra of the two models validates our assumption that the out-of-plane
flexural modes play a key role in interfacial thermal transport. Our
stack model provides clear insights into how heat dissipation from
the 2DLM is affected by the anisotropy and elastic properties of a
layered substrate like h-BN.
\begin{acknowledgments}
We gratefully acknowledge support from the Science and Engineering
Research Council through grant (152-70-00017) and use of computing
resources at the A{*}STAR Computational Resource Centre and National
Supercomputer Centre, Singapore.
\end{acknowledgments}

\bibliographystyle{apsrev4-1}
\bibliography{PaperReferences}

\end{document}